\documentstyle[proceedings,numreferences]{crckapb}
\input{psfig}

\def\Journal#1#2#3#4{(#4), {#1} {\bf #2}, #3}

\def\PRL{\em Phys. Rev. Lett.}

\def\MNRAS{{\em Mon. Not. R. Astron. Soc.}}
\def\APJ{{\em Astrop. J.}}
\def\AA{{\em Astron. Astrop.}}
\def\NAT{{\em Nature}}
\def\ASTRPH{{\em Astro-ph/}}
\def\ARAA{{\em Annu. Rev. Astron. Astrop.}}
\def\APJL{{\em Astrop. J. Lett.}}

\def\be{\begin{equation}}
\def\ee{\end{equation}}
\def\bea{\begin{eqnarray}}
\def\eea{\end{eqnarray}}

\begin{opening}

\title{ASTROPHYSICAL EVIDENCE FOR BLACK HOLE EVENT HORIZONS}

\author{ K. MENOU, E. QUATAERT AND R. NARAYAN}

\institute{Harvard-Smithsonian Center for Astrophysics, \\ 60 Garden Street,
Cambridge, MA 02138, USA}

\end{opening}

\runningtitle{EVIDENCE FOR BLACK HOLE EVENT HORIZONS}

\begin{document}

\begin{abstract}
\footnote{To appear also in the Proceedings of 
The Eighth Marcel Grossmann Meeting.}

Astronomers have discovered many potential
black holes in X-ray binaries and galactic nuclei. These black holes 
are usually identified by the fact that they are
too massive to be neutron stars. Until recently, however, there was no
convincing evidence that the objects identified as black hole
candidates actually have event horizons. This has changed with 
extensive applications of a class of accretion models for describing the
flow of gas onto compact objects; for these solutions, called
advection-dominated accretion flows (ADAFs), the black hole nature of the
accreting star, specifically its event horizon, plays an important
role.  We review the evidence that, at low luminosities, accreting
black holes in both X-ray binaries and galactic nuclei contain ADAFs
rather than the standard thin accretion disk.
\end{abstract}
  
\section{Introduction}

Astronomers have identified potential black holes in a variety of
astrophysical objects. In binary star systems, consisting of two stars
revolving around each other under the influence of their mutual
gravitational attraction, black holes with masses of order several solar 
masses ($M \sim 5-20 M_{\odot}$) are thought to exist as the dead remnant
of the initially more massive of the two
stars.

Occasionally, the black hole in such a system accretes (receives) 
matter from
the outer layers of its companion and the binary becomes a
powerful emitter of high energy radiation (hence the name: X-ray
binary, XRB) \cite{bookxrb}.
The radiation is ultimately due to the conversion of
gravitational potential energy: matter heats up as it falls into the
deep potential well of the black hole and the hot gas radiates \cite{bookfkr}.

In addition, supermassive black holes ($M \sim 10^{6}-10^{10} M_{\odot}$)
 probably exist at the centers of most galaxies. By
accreting matter (stars or gas) in their vicinity, these black holes may 
produce the
intense emission that we observe from quasars and other Active Galactic Nuclei
(AGN) \cite{bookagn}.

Since the discovery of black hole solutions in Einstein's theory of
general relativity \cite{oppsny39},\cite{ker63}, no
definitive proof of their existence has been found.  The main issue is
that a black hole remains, by its very nature, concealed to the
electromagnetic observer and can only be observed through its
indirect gravitational influence on the world beyond its event
horizon \cite{bookgrav}. 
Mass estimates of compact objects in some systems argue for the
existence of black holes. 
In addition, several observational 
signatures have been
proposed as suggestive of a black hole environment. 
These lines of evidence do not, however, constitute 
definitive proof of the reality of black holes.  

Recently, a class of accretion solutions called advection-dominated
accretion flows (ADAFs) has been extensively applied to a number of accretion
systems; for these models, most of the gravitational energy released
in the infalling gas is carried (or advected) 
with the accretion flow as thermal energy, rather than
being radiated as in standard solutions \cite{naryi94}. Since a large amount of
energy falls onto the central object, there is an
important distinction between compact objects with an event horizon
and those without; for those with an event horizon, the energy is
``lost'' to the outside observer after it falls onto the central object, 
while for those with a hard surface, 
the energy is re-radiated 
and ultimately observed \cite{naryi95b}. 
This new and potentially powerful method of discriminating between 
black holes and other kinds of compact objects is the subject of this review.

In \S 2, we list the best black hole candidates, and briefly review
the classical techniques and theories used to detect
potential black holes. In \S 3, we describe various solutions to the
astrophysical problem of gas accreting onto a central
object, emphasizing the specific characteristics of ADAFs.  In \S 4,
we describe how ADAF models have been applied to X-ray binaries and
galactic nuclei and we explain why these studies strengthen the evidence for
the existence of black holes.

\section{``Classical'' Evidence for Black Holes}
 
\subsection{Evidence based on Mass in X-Ray Binaries}
It is well known that neutron stars possess a maximum mass above which
they collapse to a black hole. The precise mass limit is not known
due to uncertainties in the physics of neutron star interiors; like white
dwarfs, neutron stars are supported by degeneracy pressure \cite{cha31}, 
but unlike white dwarfs, nuclear forces play an
important role in their structure \cite{har78}. 
The equation of state of matter at ultra-nuclear densities
is not well known. The maximum mass for the stiffest
equation of state is $\sim 3$ $M_{\odot}$ \cite{kalbay96},
and the limit goes up by no more than 25 \% if one includes the effects
of rotation \cite{friips87},\cite{bookshateu}. 
Even without detailed knowledge of the
equation of state, it is possible from first principles to put a
limit on the maximum mass. With an equation of state satisfying
causality (sound speed less than the speed of light) and assuming that
general relativity is correct, the limit is $\sim 3.4$
$M_{\odot}$ \cite{rhoruf74}. Without causality, it reaches $5.2$
$M_{\odot}$ \cite{harsab77},\cite{bookshateu}. Although
exotic compact stars with very large masses could theoretically 
exist \cite{bahlyn90},\cite{milsha97}, 
it is generally accepted by the astronomical community
that compact stellar objects with masses above $\sim 3$ $M_{\odot}$
are strong black hole candidates.

In various X-ray binary systems, one can constrain the mass
of the central object by observations of the spectral lines of the
secondary star (usually a main-sequence star).  The Doppler shifts of
these spectral lines give an estimate of the radial velocity of the
secondary relative to the observer. When combined with the orbital period 
of the binary, this can be translated (through Kepler's 3rd law) into a
``mass function'' for the compact object,
\be
f(M_X)=\frac{M_X \sin ^3 i}{(1+q)^2}. 
\ee 
The mass function does not directly give the mass $M_X$ of the compact star
because of its dependence on the unknown 
inclination $i$ between the normal to the binary orbital plane and 
the line of sight, and the
ratio of the secondary mass to the compact star mass, $q=M_2/M_X$.  
However, the mass function is a firm lower 
limit to $M_X$.  Therefore, mass functions above
$3$ $M_{\odot}$ suggest the presence of black holes.  Additional
observational data (absence or presence of eclipses, for instance, or
information on the nature of the secondary star) can
help to constrain $i$ or $q$, so that a likely value of $M_X$ can often be
determined. The best stellar mass black hole candidates currently known are
summarized in Table~\ref{tab:bhc}.  A detailed account of 
dynamical mass estimates can be found in {\it X-ray
Binaries} \cite{bookxrb}.

\begin{table}[t]
\caption{The best black hole candidates among XRBs [20],[133].
Type refers to
Low Mass X-ray Binary (relatively low mass companion star)
vs. High Mass X-ray Binary. Mass functions $f(M_X)$ and likely masses $M_X$ 
 are given in solar units ($M_{\odot}$).\label{tab:bhc}}
\vspace{0.4cm}
\begin{center}
\begin{tabular}{c c c c c }
\hline
& & & & \\
Object & Type & $f(M_X)$ & Likely $M_X$ & References \\ 
& & & & \\ \hline
& & & & \\
GRO J0422+32 (XN Per 92)& LMXB & $1.21 \pm 0.06$ & $\geq 9$ & 
\cite{beesha97},\cite{casetal95},\cite{filmat95}\\
A 0620-00 (XN Mon 75)& LMXB & $2.91 \pm 0.08$ & $4.9-10$ & 
\cite{macrem86},\cite{shanay94}\\
GRS 1124-683 (XN Mus 91) & LMXB & $3.01 \pm 0.15$ & $5-7.5$ & 
\cite{oroetal96}\\
4U 1543-47 & LMXB & $0.22 \pm 0.02$ & $1.2-7.9$ &\cite{orojai97} \\
GRO J1655-40 (XN Sco 94)& LMXB & $3.16 \pm 0.15$ & $7.02 \pm 0.22$ & 
\cite{baietal95},\cite{orobai97}\\
H 1705-250 (XN Oph 77)& LMXB & $4 \pm 0.8$ & $4.9 \pm 1.3$ & 
\cite{remetal96}\\
GS 2000+250  (XN Vul 88)& LMXB & $4.97 \pm 0.1$ & $8.5 \pm 1.5$ & 
\cite{calgar96}\\
GS 2023+338 (V404 Cyg) & LMXB & $6.08 \pm 0.06$ & $12.3 \pm 0.3$ & 
\cite{shaetal94},\cite{sanetal96}\\
& & & & \\ \hline
& & & & \\
0538-641 (LMC-X-3)    & HMXB & $2.3 \pm 0.3$ & $7-14$ & 
\cite{cowetal83}\\
1956+350 (Cyg X-1)    & HMXB & $0.24 \pm 0.01$ & $7-20$ & 
\cite{giebol82}\\
& & & & \\ \hline
\end{tabular}
\end{center}
\end{table}

\subsection{Evidence based on Mass and Velocity Profiles of Galactic Centers}
Lynden-Bell's conjecture \cite{lyn69} that AGN are powered by
accreting supermassive black holes is supported by the intense and
energetic emission seen in these systems (implying a high efficiency process
like accretion onto a compact object, cf. {\it 3.1}). 
Observations of superluminal jets in some sources (implying a 
relativistic source) confirm the hypothesis.  
In addition, rapid variability of the
luminosity, in particular the X-ray luminosity, is often
observed.  For a time-scale of variability $\Delta t$, causality
implies a source size less than $R \sim c \Delta t$, where $c$ is the speed
of light.  The estimates of $R$ from the observed variability are
typically small, thus supporting the black hole
hypothesis \cite{bookagn}. 

In recent years, new lines of evidence have emerged, thanks
to improved observational techniques. Detailed spectroscopic studies
of the central regions of nearby galaxies provide valuable
information on the line-of-sight velocities of matter (gas or stars). 
Since the presence of a
massive black hole influences the dynamics of
matter orbiting around it \cite{bookbintre}, these studies can help
discriminate between black hole and non black hole models.
Current results favor
the presence of black holes, but alternative scenarios still remain
possible \cite{korric95}. In Table~\ref{tab:smbhc} we list the best
supermassive black hole candidates revealed by these 
techniques \cite{ric97},\cite{fortsv97}.

\begin{table}[t]
\caption{Summary of supermassive black hole candidates 
at the centers of nearby galaxies. The first three are the best candidates. 
 The likely mass is given in solar units ($M_{\odot}$).\label{tab:smbhc}}
\vspace{0.4cm}
\begin{center}
\begin{tabular}{c c c}
\hline
 & & \\
Galaxy & Likely Mass & References \\
& & \\ \hline
& & \\
Milky Way (Sgr A*) & $2.5 \times 10^6 $ & \cite{eckgen97},\cite{halrie96a},\cite{halrie96b},\cite{geneck97}\\
NGC 4486 (M87) & $3.2 \times 10^9$ & \cite{m87},\cite{m87new}\\
NGC 4258 & $3.6 \times 10^7$ & \cite{ngc4258}\\
& &\\
\hline
& & \\
NGC 224 (M31) & $7.5 \times 10^7$ & \cite{korric95},\cite{fortsv97}\\
NGC 221 (M32) & $3 \times 10^6$ & \cite{m32},\cite{vandez97},\cite{vancre97}\\
NGC 1068 & $10^7$& \cite{gregwi96}\\
NGC 3115 & $2 \times 10^9$ & \cite{ngc3115}\\
NGC 3377 & $8 \times 10 ^7$ & \cite{korric95}\\
NGC 3379 & $ 0.5-2 \times 10^8 $ & \cite{ngc3379}\\
NGC 4261 & $ 4.9 \times 10^8$ & \cite{ngc4261}\\
NGC 4342 & $3.2 \times 10^8$& \cite{fortsv97} \\
NGC 4374 (M84)& $1.5 \times 10^9$ & \cite{bowgre97}\\
NGC 4486B & $6 \times 10^8$ & \cite{ngc4486b}\\
NGC 4594 & $10^9$ & \cite{ngc4594}\\
NGC 4945 & $10^6$ & \cite{gremor97}\\
NGC 6251 & $7.5 \times 10^8$ &\cite{fortsv97}\\
& & \\ 
\hline
& & \\
MCG-6-30-15& ?& \cite{tannan95},\cite{fabnan95}\\
& & \\ 
\hline
\end{tabular}
\end{center}
\end{table}

\subsection{Evidence based on X-ray Spectral Lines}
X-ray emission lines are observed in many Seyfert 1 galaxies (a
subclass of AGN). The lines are probably due to fluorescence by 
iron atoms in a relatively cold gas
irradiated by X-ray photons
\cite{geofab91},\cite{matfab97}. 
As matter nears a black
hole, the velocity of the gas becomes large. Doppler shifts, as well as
gravitational redshifts, are expected in the observed radiation. In the galaxy
MCG-6-30-15, a very broad emission line has been 
observed and is believed to originate from gas rotating
close to the central mass at a speed comparable to the speed of
light \cite{tannan95},\cite{fabnan95}. 
It has been claimed that
the Kerr rotation parameter of the postulated black hole in MCG-6-30-15 can
be constrained to
a value close to unity \cite{dabfab97}, though this has since been
challenged \cite{reybeg97}.

\subsection{Better Evidence ?}
Due to the nature of a black hole, it is difficult to prove its
existence directly.  Arguably, the best proof would be based on strong
field general relativistic effects. However, quite generally, these
effects can be well imitated by a neutron star or a 
more exotic compact
object with a radius of a few Schwarzschild radii. 
The exception is the event horizon, which cannot be mimicked
by any other object, since it is constitutive of black holes.
Advection-dominated accretion flows may provide unique evidence for
this intrinsic feature of black holes \cite{naryi95b},\cite{narmac96},\cite{narbar97},\cite{narmah97}.

\section{Accretion solutions}

\subsection{Why Accretion ?}
Astrophysics deals with a variety of sources that show either persistent
or intermittent high energy (typically X-ray) emission. Many of these sources
involve enormous amounts of released energy. Accretion onto a neutron
star or  a black hole is the most efficient known process for
converting matter into energy (excluding the obvious
matter-antimatter annihilation process), far more efficient than the
fusion of hydrogen into helium \cite{bookfkr},\cite{bookagn}.
Therefore, one is naturally inclined to explain high-energy
astrophysical observations through the accretion paradigm.  

In these models, the accreting gas radiates (in part or totally) the
energy that it acquires in falling into the gravitational potential
well of the central object.  For a given accretion rate $\dot{M}$
(quantity of matter per unit time), the gravitational energy available 
during accretion onto a spherical object of radius $R_X$ and
mass $M_X$ is 
\be L_{acc}=\frac{GM_X\dot{M}}{R_X},
\label{eq:lacc}
\ee     
where $G$ is the gravitational constant. 
The radiative efficiency of an accretion flow is defined as 
its ability to convert the 
rest mass of particles at infinity into radiated energy, 
\be
\eta=\frac{L_{rad}}{\dot{M}c^2},
\label{eq:eff}
\ee where $L_{rad}$ is the luminosity radiated by the flow (less than
$L_{acc}$) and $c$ is the speed of light.  
For a compact object, i.e. an object with a large ratio $M_X/R_X$, we see from
Eq.~\ref{eq:lacc} that the efficiency is potentially very high. 
For a neutron star 
or a black hole, $\eta$ can be $\sim 0.1$ (in contrast to the $\sim 0.007$
efficiency of hydrogen fusion).

It is convenient to scale the luminosity and the mass accretion rate in terms
of the so-called Eddington limit. The Eddington luminosity is that 
luminosity at which the radiation pressure on
a spherical flow balances the gravitational attraction of the central object: 
\be L_{Edd} = 1.3 \times 10^{38}
\left( \frac{M_X}{M_{\odot}} \right) \ {\rm erg \ s^{-1}.}
\label{eq:edd}
\ee
This is the theoretical upper limit to the luminosity
of a steady spherical accretion flow (but it is usually meaningful for
other geometrical configurations as well).
Another useful quantity is the Eddington accretion rate, defined here as the 
accretion rate at which a $0.1$ efficiency accretion flow emits the Eddington
luminosity:
\be
\dot{M}_{Edd}=\frac{L_{Edd}}{0.1  c^2}=1.4 \times 10^{18}
\left( \frac{M_X}{M_{\odot}} \right) \  {\rm g  \ s^{-1}.}
\label{eq:mdedd}
\ee
We use the following notation for the scaled accretion rate: 
\be
\dot{m}=\frac{\dot{M}}{\dot{M}_{Edd}}.
\ee

\subsection{Standard Solutions}
The equations describing the accretion of gas onto a central object are
the standard conservation laws of mass, momentum and energy, coupled with
detailed equations for the microscopic radiation processes. 
These equations are highly
nonlinear, and allow multiple solutions.  One is generally
interested only in sufficiently stable solutions, since 
these are the ones that can
be observed in nature.

Historically, the first solution to be considered
was Bondi spherical accretion \cite{bon52}, in which
matter without angular momentum accretes radially onto the central
object. This solution is usually not relevant close to the central
object (where the bulk of the observed radiation originates), since accreting
matter in astrophysical settings invariably possesses a 
finite amount of angular momentum and cannot
adopt a purely radial accretion configuration.

The second well known accretion solution is the ``thin disk'' solution
in which matter revolves around the central object in nearly Keplerian
orbits as it slowly accretes
radially \cite{shasun73},\cite{novtho73},\cite{pri81}.
To reach the central object, 
the accreting matter must get rid of its angular momentum
through viscous transport processes.  Associated with the
transport of angular momentum, viscous dissipation heats the accreting
gas, and controls, along with the cooling processes, the energetics of
the accretion flow.
In the thin disk solution,
the viscously dissipated gravitational energy is radiated very
efficiently.  In terms of the efficiency defined in Eq.~\ref{eq:eff},
a thin disk has $\eta \sim 0.06$ for a Schwarzschild black hole,
and $\eta \sim 0.42$ for a maximally rotating Kerr black
hole \cite{earpre75}. (In the case of a neutron star star, $\eta \sim 0.2$.)
Since the matter cools efficiently, it adopts a
vertically thin configuration where $H/R \ll 1$, $R$ being the distance
from the central object and $H$ the height of the disk as measured
from the orbital plane.  
The thin disk solution has been the common paradigm in
accretion theory for many years, and has been successfully applied to
disks around young stars, disks in binaries and AGN disks \cite{bookfkr}. 

The thin disk model assumes that, in steady-state (constant 
$\dot{m}$ through the disk), the local viscous dissipation rate
is everywhere exactly balanced by the locally emitted flux. 
The disk is
optically thick, which means that each annulus of the disk emits
a blackbody spectrum (to first approximation). The overall emission from
the disk is the superposition of the emission from all annuli, which means
a superposition of blackbody spectra at different temperatures.
However, since the emission is dominated by the innermost annuli, what 
emerges is similar to a single temperature blackbody spectrum. 
For a detailed review of thin accretion disk theory and its
applications, see {\it Accretion Power in Astrophysics} \cite{bookfkr}.

The thin disk model has serious difficulties
explaining the properties of some observed systems. 
In particular, it cannot account for the fact that radiation from
many systems often spans the entire electromagnetic spectrum,
with significant emission
from radio to gamma rays; this is very different from a simple blackbody
type spectrum.  

\subsection{Advection-Dominated Accretion Flows}
The thin disk paradigm describes the structure of an accretion flow
when the gas can (locally) radiate all of the viscously generated
energy. When the gas does not radiate efficiently, 
the viscously generated energy is
stored as thermal energy and advected by the flow. In this case, one has 
an ADAF, whose 
structure and properties are markedly
different from a thin disk. 

ADAFs can exist in two regimes. The first is when the gas is optically
thin and the cooling time exceeds the inflow time (defined as the 
time-scale for
matter to reach the central object) \cite{ich77},\cite{reebeg82},\cite{naryi94},\cite{naryi95b},\cite{abrche95}.
The second is when the gas is so
optically thick that a typical photon diffusion time out of the flow
exceeds the inflow time \cite{kat77},\cite{beg78},\cite{abrcze88}.
In both situations, the accretion flow is unable to cool 
efficiently and advects a significant fraction of the internally
dissipated energy \cite{cheabr95}.
We focus on optically thin ADAFs,
since they have seen the most applications to observed systems (see 
Narayan \cite{nar97} and Narayan, Mahadevan \& Quataert \cite{narmah97b} 
for reviews of ADAFs and their 
applications). 

\subsubsection{Optically thin, two-temperature ADAFs: Basic Assumptions}
Important studies of two-temperature accretion flows and advection-dominated 
accretion flows have been carried out in the 
past \cite{shalig76},\cite{ich77},\cite{reebeg82},
but a detailed and consistent picture has only emerged 
recently \cite{naryi94},\cite{naryi95a},\cite{naryi95b},\cite{abrche95},
offering the opportunity to apply specific models to
astrophysical systems.

A standard assumption in the theory of hot accretion flows is that the
gas is two temperature, with the ions significantly hotter than the
electrons.
In two-temperature ADAFs (2-T ADAFs), the viscous dissipation is
assumed to preferentially heat the ions.  The fraction of the viscous
heating that goes directly to the electrons is parameterized by a factor
$\delta$, usually set to $10^{-3}$ ($\sim m_e/m_p$).  
In addition, the electrons receive energy from the ions via Coulomb 
collisions. When the density is low, however, this process is not very 
efficient. Therefore, only a small fraction of the viscous energy reaches 
the electrons; this energy is usually radiated (though not 
always \cite{nakkus97},\cite{mahqua97}). 
The rest of the energy remains in the ions, which are unable   
to cool in an inflow time. The energy in the ions is thus advected 
with the flow and
is finally deposited on the central object.  If the object is a black hole, the
energy disappears; if it has a surface,
the advected energy is re-radiated. 

The efficiency of the turbulent viscous transport and dissipation in
the flow are parameterized, as in the standard thin disk theory, by
the ``$\alpha$-prescription'' \cite{shasun73}: 
the viscosity is written as $\nu= \alpha
c_s^2 / \Omega_K$, where $\alpha$ is the viscosity parameter, $c_s$ is
the sound speed in the plasma and $\Omega_K$ is the local Keplerian
angular velocity. If we assume that the Balbus-Hawley 
instability \cite{balhaw91} is
the origin of the MHD turbulence in the flow, we expect the magnetic
fields and the gas in an ADAF to be in equipartition (i.e. magnetic
pressure comparable to gas pressure), and the viscosity parameter $\alpha$ to
be close to $0.3$ \cite{hawgam96}.

\subsubsection{2-T ADAFs: Properties of the Flow and the Emitted Radiation}  
The dynamical properties of ADAFs were first described
by analytical self-similar solutions \cite{naryi94}; these were soon
replaced by global numerical
solutions \cite{abrche96},\cite{narkat97},\cite{cheabr97},\cite{gampop97},\cite{popgam97},\cite{peiapp97}.
The flows are geometrically thick ($H/R \sim 1$) and are well
approximated as spherical flows with most of the gas properties
roughly constant on spherical shells \cite{naryi95a}. Matter falls
onto the central object at radial speeds less than, but comparable to,
the free fall velocity and with angular velocities significantly less
than the local Keplerian value. ADAFs are radially convectively
unstable, but the implications of this for the structure and dynamics
of the flows are still unclear \cite{naryi95b}.

A self-consistent determination of the thermal structure of an ADAF
requires detailed numerical
calculations \cite{naryi95b},\cite{narbar97}. The poor
energetic coupling between ions and electrons, and the advection of
heat by the ions, is reflected in their temperature profiles: the ion
temperature goes as $T_i \sim 10^{12}K/r$  ($r$ is the distance from the
central object in units of the Schwarzschild radius) while the electron
temperature saturates at $T_e \sim 10^{9-10}$ K in the inner
$10^2-10^3$ Schwarzschild radii. 
Above a critical accretion rate,
$\dot{m}_{crit}$, the density in the plasma becomes sufficiently
high that Coulomb
collisions efficiently transfer energy from the ions to the electrons.
The gas then radiates most of the viscously dissipated energy and is no longer 
advection-dominated. 
ADAFs therefore only exist for accretion rates below $\dot{m}_{crit}
\sim \alpha^2$ \cite{ich77},\cite{reebeg82},\cite{naryi95b},\cite{abrche95}. 
For $\dot{m} > \dot{m}_{crit}$, accretion occurs as a cool thin disk.

Since ADAFs are optically thin and most of the viscously released energy is
advected, they are significantly underluminous
compared to a thin accretion disk with the same accretion rate; in fact, 
the radiative efficiency can be as low as $\eta \sim 10^{-3}-10^{-4}$ at 
low $\dot{m}$.
The emission by the hot plasma comes almost entirely from the
electrons.  Bremsstrahlung is an important emission process. In addition, since
the electrons are in the presence of significant magnetic fields and
are marginally relativistic, 
synchrotron emission and inverse Compton
radiation are also very important.  Synchrotron emission is the
dominant mechanism at radio, infrared and optical
wavelengths, with the peak emission occurring at a wavelength that depends
on the black hole mass ($\lambda_{peak} \propto M_X^{1/2}$) \cite{mah97}. 
Comptonization of soft photons by hot electrons
contributes from infrared to hard X-rays and is a strong function 
of the accretion rate.  Bremsstrahlung emission generally
dominates in the X-ray band at low mass accretion rates, but
Comptonization takes over at higher accretion rates. 
A significant amount of $\gamma$-rays
is also emitted, resulting from neutral pions (created by
proton-proton collisions) which decay into very hard
photons \cite{mahnar97}. Pair creation processes are not
very important in ADAFs since the radiation energy densities in the flow are
low \cite{bjoabr96},\cite{kusmin96}.

\subsubsection{2-T ADAFs : Stability}
The stability of accretion solutions is an important issue in
accretion physics, since an accretion solution is
not viable if it is unstable (by that one generally means
linearly unstable). For instance, the important 2-T (non advective)
solution of Shapiro et al. \cite{shalig76} is known to be violently
unstable \cite{pir78} and for this reason will probably never be
observed in nature. The stability of 2-T ADAF solutions has been
investigated in the long and short wavelength
limits \cite{naryi95b},\cite{mantak96},\cite{katyam97},\cite{wu97a},\cite{wu97b}. 
These studies conclude that, 
since the only unstable modes have sufficiently slow growth rates, ADAFs 
constitute a viable solution.
 
\section{Applications of ADAFs}
ADAF models have been applied to a number of low luminosity systems.
They give a satisfying description of the spectral characteristics of
the source Sgr A* at the center of our Milky Way 
Galaxy \cite{narmah97},\cite{manmin97}, 
the weak AGN NGC 4258 \cite{lasabr96} and M87 \cite{reydim96}, and
of several quiescent black hole 
XRBs \cite{narmac96},\cite{narbar97},\cite{hamlas97}.
All of these systems are known to
experience low efficiency accretion and the thin disk solution
encounters serious difficulties in explaining the observed spectral properties.
In addition, the ADAF model also fits brighter systems with higher 
efficiencies \cite{nar96},\cite{esimac97},\cite{esietal97}.

It is worth mentioning here that previous studies have shown that
the spectra of some of these systems can be explained by constructing
models of optically thin accreting plasmas. However, these models are
not necessarily dynamically consistent and do not explicitly attempt 
to satisfy
mass, angular momentum and energy conservation.  The ADAF
solutions solve the observational and theoretical problems, for the first 
time, in a reliable and dynamically consistent way.

\subsection{Modeling Techniques}
Popham \& Gammie \cite{popgam97}, among others \cite{abrche96},\cite{peiapp97},have computed the steady-state 
dynamical structure of ADAFs in the
Kerr geometry. Their numerical solutions have been used in the spectral models
presented here. 
The dynamical model provides the radial profiles of quantities such as radial speed, 
angular speed, density, pressure and viscous dissipation. 
In these models, the structure is 
determined by 3 
parameters : $\alpha$ (viscosity 
parameter), $\gamma$ (adiabatic index of the gas), and $f$ (advection 
parameter, i.e. the fraction of the viscously dissipated energy which is 
advected). Usually, models assume that a constant fraction 
$(1-\beta)=0.5$ of the 
total pressure comes from the magnetic field pressure and that 
$\alpha \sim 0.6(1-\beta) =0.3$. 
The adiabatic index is given by 
$\gamma=(8-3 \beta)/(6-3 \beta)=1.44$ \cite{esi97}.
The parameter $f$ is solved self-consistently by calculating the radiation
processes in detail and feeding back the information to the dynamical 
solution \cite{esimac97}. 
So far, studies have been restricted to the
Schwarzschild geometry, but ultimately, in full generality, 
the Kerr rotation parameter $a$ of the black hole will 
constitute an additional parameter.

\begin{figure}
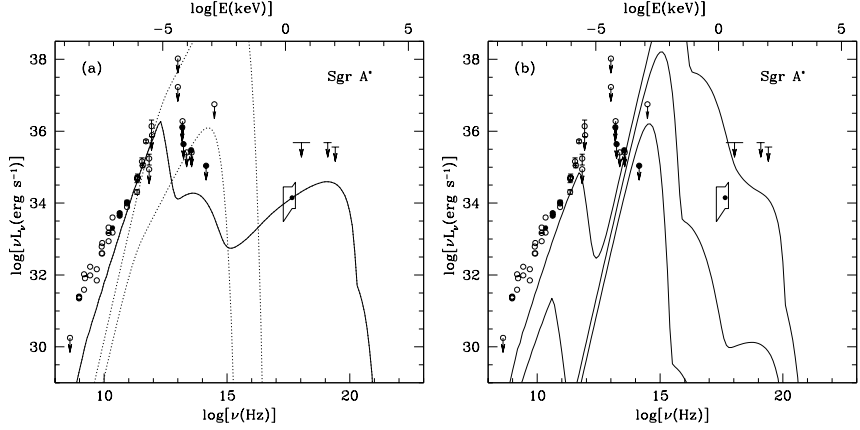

\begin{center}
\begin{minipage}[t]{0.45\hsize}
\begin{displaymath}
\psfig{figure=sgra.epsi,height=2.3in}
\end{displaymath}
\end{minipage}
\begin{minipage}[t]{0.45\hsize}
\begin{displaymath}
\psfig{figure=sgraalt.epsi,height=2.3in}
\end{displaymath}
\end{minipage}
\end{center}
\caption{(a) Spectrum of an ADAF model of Sgr A* (solid line, based on  
[92]).
The mass accretion rate
inferred from this model is $\dot{m} = 1.3 \times 10^{-4}$ in Eddington units,
in agreement with an independent observational determination.
Dotted lines show the spectra of thin accretion disks,  at the same
accretion rate (upper) and at $\dot{m} = 1 \times 10^{-8}$ (lower).
Observational data
are shown as circles with error bars, and upper limits are indicated
by arrows. The box indicates the constraints on the flux and the spectral
index in soft X-rays.
(b) Spectra of ADAF models of Sgr A* where the central mass is taken to have
a surface at $3$ $R_{\rm Schw}$ and the advected energy is assumed to be
re-radiated as a blackbody. From top to bottom, the three spectra correspond to
$\dot{m}=10^{-4},10^{-6},10^{-8}$. All three models violate the infrared limit.
}
\label{fig:sgra}
\end{figure}

\begin{figure}
\begin{displaymath}
\psfig{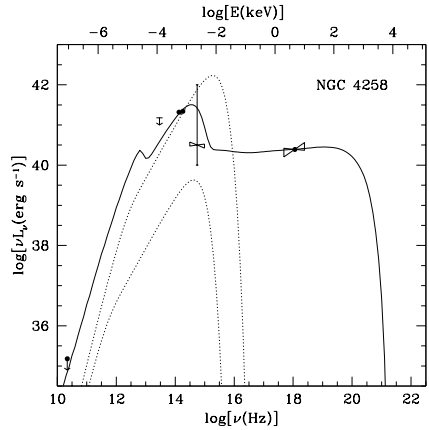}
\end{displaymath}
\caption{Spectrum of an ADAF model of NGC 4258 (solid line), at an accretion
rate $\dot{m} = 9\times 10^{-3}$, with a transition radius 
$R_{trans}=30 R_{\rm Schw}$.
Dotted lines show the spectra of thin accretion disks,  at an
accretion rate $\dot{m} = 4 \times 10^{-3}$ (upper, adjusted to fit the 
infrared points) and $\dot{m} = 10^{-5}$ (lower).
\label{fig:ngc4258}}
\end{figure}

\begin{figure}
\begin{displaymath}
\psfig{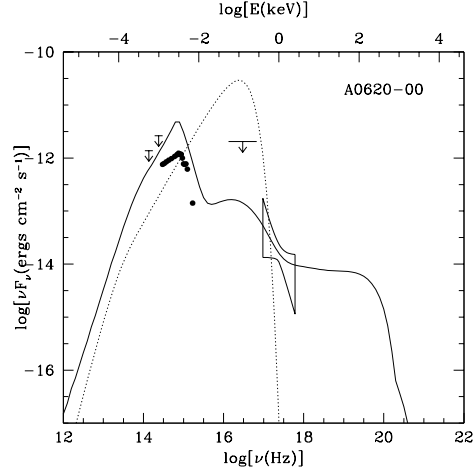}
\end{displaymath}
\caption{Spectrum of an ADAF model of A0620-00 (solid line, based on  
[88])
at an accretion rate $\dot{m}=4 \times 10^{-4}$, 
compared with the observational data.
The dotted line shows the spectrum of a thin accretion disk
with an accretion rate $\dot{m} = 10^{-5}$ (adjusted to fit the optical flux). 
\label{fig:a0620}}
\end{figure}

\begin{figure}
\begin{displaymath}
\psfig{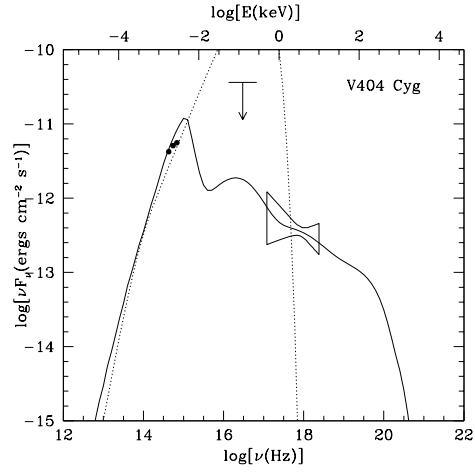}
\end{displaymath}
\caption{Spectrum of an ADAF model of V404 Cyg (solid line, based on ref.
[88])
at an accretion rate $\dot{m}=2 \times 10^{-3}$,
compared with the observational data.
 The dotted line shows the spectrum of a thin accretion disk 
with $\dot{m} = 1.8 \times 10^{-3}$ (adjusted to fit the
optical flux).
\label{fig:v404}
}
\end{figure}

\begin{figure}
\begin{displaymath}
\psfig{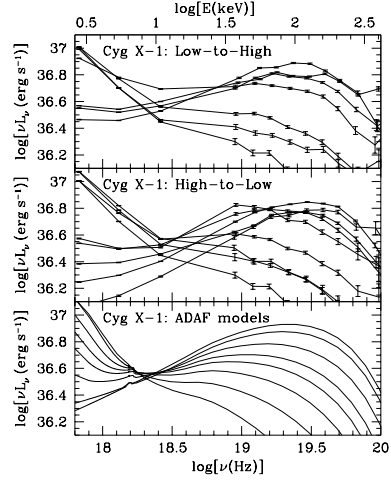}
\end{displaymath}
\caption{The broadband simultaneous RXTE (1.3-12 keV) and BATSE (20-600 keV)
spectra of Cyg X-1 observed during the 1996 low-high (upper panel) and
high-low (middle panel) state transitions. 
The bottom panel shows a sequence of ADAF models which are in good agreement
with the observations [29].}
\label{fig:cygx1}
\end{figure}

\begin{figure}
\begin{displaymath}
\psfig{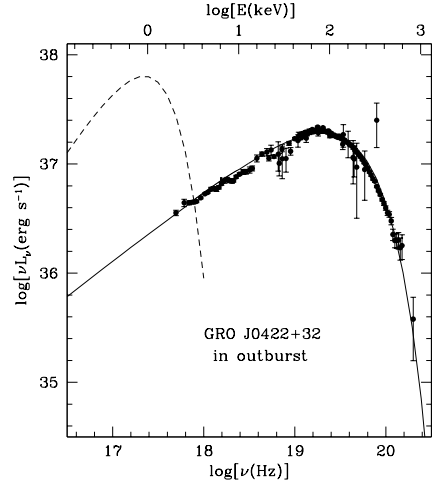}
\end{displaymath}
\caption{An ADAF model of J0422+32 (solid line) in
the so-called low state (which is actually during outburst),
compared with the observational data (dots and errorbars) [29].
The dashed line shows a thin disk model at the same accretion rate
$\dot{m}=0.1$.
\label{fig:j0422}}
\end{figure}

The steady-state energetic structure of the flow is found by solving the 
electron and proton energy equations. The emission 
spectrum is obtained as a byproduct of this procedure, in which synchrotron,
bremsstrahlung and Compton processes are all taken into account. Since soft 
photons can be scattered off electrons more than once before 
escaping the flow, the Compton problem has to be solved by a global 
``iterative scattering method'' for consistency \cite{narbar97}.
In the models presented here,
the photon transport has been treated as Newtonian, but including 
gravitational 
redshift effects. Generic ADAF spectra with photon transport in the Kerr 
metric have been computed by Jaroszynski \& Kurpiewski \cite{jarkur97}.

The basic ADAF model has one adjustable parameter: the mass 
accretion rate $\dot{m}$.
The X-ray flux emitted by the ADAF is very sensitive to the density and 
therefore the accretion rate. For this reason, 
$\dot{m}$ is usually adjusted
so that the resulting spectrum fits the available X-ray data.

The spectral characteristics of some systems are better explained by
mixed ADAF + thin disk models. The accretion proceeds via
an ADAF in the inner regions of the accretion flow, and via a thin disk
beyond a so-called transition radius ($R_{trans}$).
The contribution of the thin disk
to the overall spectrum is a strongly decreasing function of $R_{trans}$,
so that $R_{trans}$ is a relevant parameter (and therefore is well determined)
only in those systems
where the thin disk contribution to the emitted flux is significant (see 
NGC $4258$ and Cyg X-1, for instance).

ADAF models and the corresponding observational data points and constraints 
are shown in Fig.~\ref{fig:sgra} to Fig.~\ref{fig:j0422} for a 
number of well-known black hole systems.
Spectra of thin accretion disks are 
also shown. While the ADAF models are in qualitative agreement with the 
observations, the thin disk models are generally ruled out quite convincingly.

\subsection{Spectral Models of Low Luminosity Systems}

\subsubsection{Sgr A*}    
Sgr A* is an enigmatic radio source at the center of our Galaxy. 
 Dynamical mass estimates show that a dark mass $M \sim 2.5 \times 10^6 M_{\odot}$ resides in a region of less than 0.1 pc in extension around this 
source \cite{eckgen97},\cite{halrie96a},\cite{halrie96b}. 
Based on this, it is assumed that Sgr A* is a supermassive black hole.
Observations of gas flows and stellar winds in its vicinity suggest that Sgr A* accretes matter at a rate in the range $\dot{M} \sim$ a few $\times 10^{-6}-10^{-4} M_{\odot} \ {\rm yr^{-1}}$ ($\dot{m} \sim 
10^{-4}-10^{-2}$) \cite{genetal94},\cite{mel92}.
A standard thin disk accreting with this $\dot{m}$ at a typical efficiency $\eta \sim 0.1$ is far too luminous to account for the observed integrated flux. Such a model also disagrees with the observed shape of the spectrum  and is ruled out (by orders of magnitude) by an upper limit on the Sgr A* flux in the near infrared (Fig.~\ref{fig:sgra}a).

The first suggestion that Sgr A* may be advecting a significant amount of energy was made 
by Rees \cite{ree82}, and the first ADAF spectral model of this source
appeared in Narayan, Yi \& 
Mahadevan \cite{naryi95}.
Since then, the modeling techniques \cite{nakmat96},\cite{nakkus97}
and observational 
constraints have improved,
 allowing the construction of a refined model of Sgr A* \cite{narmah97},\cite{manmin97}. 
The model is consistent with the 
{\it independent} estimates of $M$ and $\dot{M}$ mentioned above, 
and it fits the observed fluxes in the radio, millimeter and X-ray bands 
and upper limits in the sub-millimeter and infrared bands 
(see Fig.~\ref{fig:sgra}a). 

In evaluating the quality of the fit in Fig.~\ref{fig:sgra}a, it should be kept
in mind that only one parameter, $\dot{m}$, has been adjusted; this parameter
has been optimized to fit the X-ray flux at $2$ keV. The position of the
synchrotron peak at $10^{12}$ Hz and its amplitude are not fitted but
are predicted by the model; the agreement with the data is good. There is,
however, a problem at lower radio frequencies, $\leq 10^{10}$ Hz, where the
model is well below the observed flux. This is currently unexplained.

An important feature of the ADAF model of Sgr A* is that 
the observed low luminosity of the source
is explained as a natural consequence of the advection of energy in the flow (the radiative efficiency is very low,  
$\eta \sim 5 \times 10^{-6}$) and the disappearance 
of this energy through the event horizon of the black hole. 
The model will not work if the central object has a hard surface,
as demonstrated in Fig.~\ref{fig:sgra}b.

\subsubsection{NGC 4258}
The mass of the central black hole in the AGN NGC 4258 has been measured to be $3.6 \times 10^7 M_{\odot}$ \cite{ngc4258}. Highlighting the fact that the observed optical/UV and X-ray luminosities are significantly sub-Eddington ($\sim 10^{-4}$ and $\sim 10^{-5}$ respectively), Lasota et al. \cite{lasabr96} proposed that accretion in NGC 4258 proceeds through an ADAF. 
They found that the emission spectrum and the low luminosity of this system can be explained by an ADAF model, provided that most of the viscously dissipated energy in the flow is advected into a central black hole.
Since then, new infrared measurements have been made \cite{chabec97} 
that constrain the transition radius
to be $R_{trans} \sim 30$ Schwarzschild radii. 
The ADAF produces the X-ray emission while the outer thin disk 
accounts for the newly observed 
infrared emission (see Fig.~\ref{fig:ngc4258}, 
unpublished). The refined model is also in agreement with a revised 
upper limit on the radio flux \cite{private}.
Maoz \& McKee \cite{maomck97} and Kumar \cite{kum97} find, via quite independent arguments, 
an accretion rate for NGC 4258 in agreement with the ADAF model. Neufeld 
\& Maloney \cite{neumal95} estimate a much lower $\dot{m}$ ($\leq 10^{-5}$) in 
order to explain the maser emission in the source, but it is hard to 
explain the observed spectrum with such an $\dot{m}$ (Fig.~\ref{fig:ngc4258}).

\subsubsection{Other Low Luminosity Galactic Nuclei} 
Quasars are luminous high redshift AGN, which are believed to be
powered by supermassive
black holes with masses around $10^8-10^9$ $M_{\odot}$ \cite{ree84}. 
Most nearby
bright elliptical galaxies are believed to host dead quasars, i.e. quasars
which have become inactive
through evolution \cite{sol82},\cite{chotur92}.
As pointed out by Fabian \& Canizares \cite{fabcan88}, however, the nuclei 
of these elliptical galaxies
are much too dim, given the mass accretion rates inferred from independent
methods.
Fabian \& Rees \cite{fabree95}
suggested that the problem could be resolved if the accretion in these nuclei 
is occurring through an ADAF. 
This proposition has been confirmed
by Mahadevan \cite{mah97}
and models have been developed for specific galaxies such as 
M87 \cite{reydim96} and M60 \cite{dimfab97a}.
Lasota et al. \cite{lasabr96} have similarly argued that
low-luminosity LINER and Seyfert galaxies 
have supermassive black holes in their nuclei accreting via ADAFs. 

\subsubsection{A0620-00 and V404 Cyg}
Soft X-ray Transients (SXTs) are a class of mass transfer X-ray binaries 
in which the accreting star is often a black hole candidate. 
Episodically,  these systems enter high luminosity ``outburst'' phases, 
but for most of the time they remain in a very low luminosity ``quiescent'' phase. One of the main issues in modeling black hole SXTs in quiescence is that a thin accretion disk cannot explain both the observed low luminosity and the X-ray flux in these systems : at an accretion rate low enough to fit the observed luminosity, a thin disk will not emit any significant flux in the X-ray band.
Moreover, the X-ray spectrum will have the wrong shape. 
Narayan, McClintock \& Yi \cite{narmac96} argued that the dilemma can be solved by means of the ADAF model and demonstrated this by carrying out spectral fits of A0620-00, V404 Cyg and Nova Mus 91 in quiescence.
Narayan, Barret \& McClintock \cite{narbar97} have 
recently refined these models for V404 Cyg and A0620-00 
(see Fig.~\ref{fig:a0620} and Fig.~\ref{fig:v404}).
The models are in good agreement with the observed shapes of the X-ray spectra
(recall that the X-ray flux level is fitted by adjusting $\dot{m}$, but the 
spectral slope is unconstrained in the fit) and they satisfy other observational constraints such as optical measurements and extreme ultraviolet upper limits.
In the optical, the models are somewhat too luminous to fit the data;
the shapes of the spectra, however, are predicted well (see especially 
Fig.~\ref{fig:a0620}, where the position of the peak is reproduced well by
the model). In contrast, the thin disk model
deviates very significantly from the data and is easily ruled out. 
Hameury et al. \cite{hamlas97} showed that observations of another SXT, 
GRO J1655-40, are consistent with the presence of an ADAF in quiescence. In
addition, their model provides a convincing explanation for the time delay
that was observed between the optical and X-ray light curves during a
recent outburst \cite{ororem97}.

\subsection{More Luminous Systems}

\subsubsection{Spectral States of XRBs}

Narayan \cite{nar96} suggested that the various spectral states (from 
quiescence to the luminous outburst phase) of black hole binaries (especially 
SXTs) can be understood in terms of a sequence of thin disk + ADAF models, 
where the thin disk accounts for most of the luminosity in outburst and is 
gradually replaced by an ADAF during the transition to quiescence. 
Recently, Esin et al. \cite{esimac97}
have developed these ideas with detailed calculations, and succeeded in 
explaining
the observed spectral states of the system Nova Mus 91 surprisingly well.
The same models also appear to explain other luminous black hole systems such 
as Cyg X-1 (Fig.~\ref{fig:cygx1}) and 
J0422+32 (in outburst, Fig.~\ref{fig:j0422}) \cite{esietal97}.
These studies, for the first time, unify a fair fraction of the phenomenology 
of black hole XRBs.
Once again, the existence of a central black hole, into which 
energy is lost, is an essential feature of the models.

\begin{figure}
\begin{displaymath}
\psfig{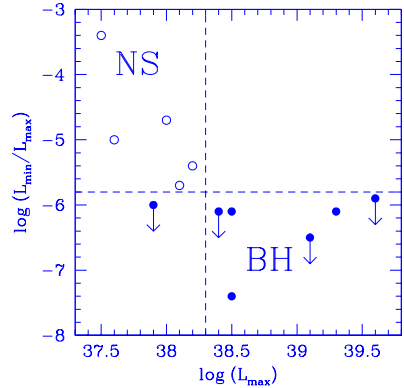}
\end{displaymath}
\caption{A comparison between black hole (BH, filled circles) and neutron 
star (NS, open circles) SXT 
luminosity variations [41].
The ratio of the quiescent luminosity to the peak outburst luminosity is
systematically smaller for BH systems than for NS systems. This  
confirms the presence of event horizons in BH candidates.
\label{fig:lumxrb}}
\end{figure}

\subsubsection{Galaxies and Quasars}
A similar unification is also possible in our
understanding of the various degrees of activity in galactic nuclei.
Yi \cite{yi96} has proposed an explanation for the apparent evolution of 
quasars from a large population of luminous objects at redshift $\geq 2$ to a 
much smaller population in the nearby universe. He postulates that 
the mass accretion rate in
galactic nuclei has decreased over cosmic time. Supermassive black holes
in galactic nuclei had $\dot{m} > \dot{m}_{crit}$ at early times and 
accreted through thin disks with a high radiative efficiency.
As $\dot{m}$ decreased with time, the accretion flow switched 
to a low efficiency ADAF. This explains the absence of very luminous 
AGN in the local neighborhood of our Galaxy.

Di Matteo \& Fabian \cite{dimfab97b} proposed that a significant fraction 
of the hard X-ray ($\geq 2$ keV) background could be due to  
emission from a population of galaxies undergoing advection-dominated accretion
in their nuclei. Yi \& Boughn \cite{yibou97} discuss a test 
of the ADAF paradigm in this context;
they argue that measurements 
of the radio and X-ray fluxes could provide direct estimates of the central 
black hole masses in many galaxies.

\subsubsection{Neutron Star SXTs vs. Black Hole SXTs}
As mentioned earlier, SXTs are XRBs that exhibit large 
luminosity variations. Most of the time they remain in a quiescent 
state, in which their luminosities are low and the transferred mass is 
partially stored in a non-steady thin disk. Episodically, they experience 
an outburst event, during which their luminosities become very 
high \cite{las96}.
During outbursts, which correspond to a sudden accretion of the 
mass stored in the disk, 
these sources reach luminosities of order the 
Eddington luminosity 
(cf. Eq.~\ref{eq:edd}). From dynamical mass estimates, and observations 
of ``X-ray bursts'' (very probably thermonuclear explosions on
the surface of neutron stars), one can distinguish between neutron star and 
black hole candidate systems \cite{bookxrb}.
Since black holes are 
more massive than neutron stars (and the Eddington luminosity scales with the 
mass of the accreting object), black hole SXTs are naturally observed in outburst at a
higher maximum luminosity than neutron star SXTs. 

Assuming that the ADAF model of SXTs in 
quiescence \cite{narmac96},\cite{narbar97} 
is generic, Narayan, Garcia \& McClintock \cite{nargar97} 
argued that black hole SXTs
 should experience more substantial luminosity changes from quiescence to 
outburst than neutron star SXTs. 
In quiescence, the central object accretes matter through an ADAF : most of 
the viscously dissipated energy is advected by the flow onto the 
central object. 
In the neutron star case, this advected energy has
 to be re-radiated from the stellar surface, whereas in the black hole case, 
it is lost through the horizon. 
Therefore, black hole SXTs in quiescence must be much less luminous 
(relative to their maximum luminosity) than neutron star SXTs, which is 
exactly what is observed in nature (see Fig.~\ref{fig:lumxrb}). 
This systematic 
effect confirms the presence of an event horizon in black hole 
SXTs \cite{nargar97},\cite{garmac97}.

\section{Conclusion}
Advection-dominated accretion flow models of black holes in various 
XRBs and AGN are in qualitative agreement with the best current 
observations. They 
provide a satisfying explanation for both the spectral characteristics 
and the extremely low luminosities
observed in these systems.
The models require that a very 
large fraction of the dissipated energy in the flow is advected into the 
central black hole. If a standard star with a surface, 
and not an event horizon, were present at the 
center of these systems, changes in the spectra and luminosities by 
orders of magnitude are predicted (e.g. compare Fig.~\ref{fig:sgra}a and
Fig.~\ref{fig:sgra}b). 
The success of ADAF models 
thus constitutes strong evidence for the existence 
of stellar mass and supermassive black holes in the Universe.

We conclude with an interesting observation.  If one considers the
candidate black hole XRBs listed in Table 1, apart from LMC X-3, all
the other systems spend the bulk of their time in the ADAF state and
only occasionally switch to the more efficient thin accretion disk
state.  Similarly, in Table 2, the three best candidates, Sgr A$^*$,
NGC 4258 and M87, appear to have ADAFs, and it is plausible that many of the
other systems have ADAFs as well, except for NGC 1068, MCG-6-30-15, 
and possibly NGC 4945.  
If ADAFs are, as the evidence suggests, very common, there is
in principle no difficulty finding systems in which to test for the presence
of event horizons.  The challenge at the moment is primarily
observational --- ADAFs by their nature are very dim and one needs
instruments with superior sensitivity to achieve adequate
signal-to-noise.

\section*{Acknowledgments}
We are grateful to Vicky Kalogera, Chris Kochanek, Avi Loeb, 
Rohan Mahadevan, Jeff McClintock and Doug Richstone for useful discussions. 
Fig.~\ref{fig:cygx1} and Fig.~\ref{fig:j0422} were kindly 
provided by Ann Esin. 
The electronic preprints referenced below can be found at the Los
Alamos National Laboratory e-print service : http://xxx.lanl.gov/.
This work was supported in part by NSF grant AST 9423209 and NASA grant
NAG 5-2837. EQ was supported by a NSF graduate research fellowship.

\end{document}